\documentclass{iopart}
\usepackage{iopams}
\begin{document}

\jl{2}

\newcommand\Bm[1]{\mbox{\boldmath$#1$}}
\newcommand\zi{\mathrm i}
\newcommand\rmc{\mathrm c}
\newcommand\rmL{\rm L}
\newcommand\rmV{\rm V}
\newcommand\rmA{\mathrm A}
\newcommand\XN{\Bm{X}_N}
\newcommand\XNp{\Bm{X}_{N+1}}
\newcommand\ilm{i\ell m}
\newcommand\lm{\ell m}
\newcommand{\lmnu}{\ell_\nu  m_{\nu}}
\newcommand{\lmmu}{\ell_\mu  m_{\mu}}
\newcommand{\lmnup}{\ell_{\nu'}m_{\nu'}}
\newcommand\no{n_{\mathrm o}}
\def\rlim{\raisebox{-2.mm}{$\;\stackrel{{\textstyle\longrightarrow}}
{{\scriptstyle r\rightarrow\infty}}\;$}}

\def\bss#1{\protect\mathbf{#1}}

\title[RMF theory of laser-assisted electron-atom scattering]
{\textit{R}-matrix Floquet theory for laser-assisted electron-atom 
scattering}

\author{M Terao-Dunseath and K M Dunseath}

\address{Laboratoire de Physique des Atomes, Lasers, Mol\'ecules et
Surfaces, UMR 6627 du CNRS -- Universit\'e de Rennes 1, Campus de 
Beaulieu, F-35042 Rennes Cedex, France}

\begin{abstract}
A new version of the $R$-matrix Floquet theory for laser-assisted
electron-atom scattering is presented. The theory is non-perturbative
and applicable to a non-relativistic many-electron atom or ion in a
homogeneous linearly polarized field. It is based on the use of
channel functions built from field-dressed target states, which
greatly simplifies the general formalism.
\end{abstract}

\submitted
\maketitle

\section{Introduction}

With the widespread availability of powerful lasers, the interaction
of atomic systems with intense electromagnetic radiation has become
the subject of much study. A variety of multiphoton phenomena have
been observed, remarkable by their non-linearity. The field may induce
new processes that can only occur through the absorption or emission
of more than one photon. Multiphoton ionization, above-threshold
ionization and harmonic generation have attracted most interest, while
electron-atom collisions in the presence of a laser field have
received comparatively less attention. These are commonly described as
laser-assisted, in that the field is not required for scattering to
occur, although its presence can give rise to new inelastic processes
such as simultaneous electron-photon excitation.

Relatively little experimental work has been devoted to electron-atom
collisions in a laser field.  A number of groups have investigated
elastic or free-free scattering on rare gas targets (usually argon) in
a CO$_2$ laser field at intensities up to 10$^9$ W/cm$^2$ (Andrick and
Langhans 1976, 1978; Weingartshofer \etal 1977, 1979, 1983; Bader
1986; Wallbank \etal 1987a,b, 1992). Measurements have been performed
over a wide range of incident electron energies, mostly at large
scattering angles corresponding to backward scattering, for
geometries in which the polarization axis is either parallel to the
momentum of the incident electron, or parallel to the direction of
momentum transfer. More recent experiments (Wallbank and Holmes 1993,
1994a,b) measured angular distributions for geometries in which the
polarization axis is almost perpendicular to the momentum
transfer. The first experimental investigation of simultaneous
electron-photon excitation was performed by Mason and Newell (1987,
1989) on helium in the field of a CW CO$_2$ laser at intensities from
10$^4$ to 10$^5$ W/cm$^2$. This was soon followed by studies at
intensities up to 10$^8$ W/cm$^2$ (Wallbank \etal 1988, 1990). The
only experiment at a higher frequency is that of Luan \etal (1991) who
used a Nd-YAG laser of intensity 10$^{10}$ W/cm$^2$. Details can be
found in the review by Mason (1993).

A larger number of theoretical studies have been undertaken, and are
comprehensively reviewed by Ehlotzky \etal (1998). The challenge for
theory lies in accurately treating both radiative couplings and the
dynamics of the collisional system over very large distances: each of
the electron-target, laser-electron and laser-target interactions
needs to be taken into account. If one of these dominates, a
perturbative approach can be adopted. In early work on free-free
scattering for example, Bunkin and Fedorov (1966) treated the
laser-electron interaction to all orders using Gordan-Volkov wave
functions, and the electron-target interaction to first order using
the Born approximation. The dressing of the target by the field was
neglected. Kroll and Watson (1973) included higher order terms in the
Born series to obtain a formula valid when the scattering potential is
weak or when the frequency of the laser field is small (the low
frequency approximation).  An alternative approach for treating the
laser-atom interaction non-perturbatively is furnished by Floquet
theory. This is applicable to periodic Hamiltonians and reduces the
problem to be solved to that of a set of time-independent coupled
equations. The Floquet expansion has been combined for example with a
partial wave expansion for potential scattering (Dimou and Faisal
1987), and with the Lippmann-Schwinger integral equation to study
low-energy free-free scattering by various potentials (Kylstra and
Joachain 1998, 1999). Most work on inelastic processes has
concentrated on high collision energies using methods based on the
Born approximation (Rahman and Faisal 1976, Jetzke \etal 1984, 1987,
Francken \etal 1988). Lower energies have been considered mainly
within the low-frequency approximation, which is valid for CO$_2$
lasers (Geltman and Maquet 1989, Mittleman 1993, Fainstein \etal 1995)
but not for Nd-YAG. An attempt to go beyond this approximation using a
pseudo-potential model has been discussed by Faisal (1987).

There now exists a variety of powerful computational methods for the
study of electron collisions with complex atoms. $R$-matrix theory for
example has been developed over many years (Burke and Berrington 1993,
Burke \etal 1994) and is now used regularly for the production of
vital atomic data needed in a variety of applications. The main
feature of this theory is the division of configuration space into
inner and outer regions, in each of which the wave function can be
expanded in a locally adapted basis. The global wave function is then
built by matching the solutions at their common boundary, using the
inverse of the log-derivative matrix or $R$-matrix. An extension of
the method to multiphoton processes was proposed by Burke \etal (1990,
1991, hereafter referred to as paper I), using the Floquet expansion
to describe the interaction of the atomic system with a laser field. A
practical solution in the outer region was developed by D\"orr \etal
(1992), hereafter referred to as paper II. Subsequently, this
$R$-matrix Floquet (RMF) theory has been used to study multiphoton
ionization of numerous atoms and ions, laser-induced degenerate states
(LIDS) and harmonic generation (see for example the review by Joachain
\etal 2000). A version describing multiphoton processes involving
diatomic molecules has recently been developed (Burke \etal 2000) and
applied to multiphoton ionization of H$_2$ (Colgan \etal 2001).

In papers I and II, the RMF theory was presented mainly with
multiphoton ionization in mind. Its application to laser-assisted
scattering has been described in detail only for the case of a
potential (electron-proton scattering, D\"orr \etal 1995). We have
since extended the method to laser-assisted scattering by atomic
targets. We first applied it to the study of free-free processes in
e$^-$--H (Charlo \etal 1998) and e$^-$--He (Charlo 1999) collisions in
a CO$_2$ laser field. We showed that the differential cross sections
display a deep minimum when the momentum transfer is nearly
perpendicular to the direction of the laser field, in accordance with
the low frequency approximation. More recently, we have studied the
simultaneous electron-photon excitation of helium in a Nd-YAG laser
field (Terao-Dunseath \etal 2001), for which the low frequency
approximation is no longer valid. We showed that the cross section is
dominated by the He$^-$(1s2s$^2$ $^2$S) resonance and that a strong AC
Stark mixing occurs between the 1s2s $^3$S and 1s2p $^3$P$^{\mathrm
o}$ states. The purpose of this paper is to present the modifications
that needed to be introduced in the original $R$-matrix Floquet theory
in order to identify unambigously field-dressed target states and to
calculate cross sections for laser-assisted scattering by a general
atomic or ionic system.

Atomic units are used throughout. We shall also use the Fano-Racah
phase convention for the spherical harmonics, with
\[Y_{\ell m}(\theta,\phi)=\zi^{\ell}Y_{\ell m}^\mathrm{CS}(\theta,\phi)\]
where CS implies the usual Condon-Shortley phase.

\section{The Floquet approach: basic equations and properties}

We consider the scattering of an electron by an atomic system composed
of an infinitely heavy nucleus of charge $Z$ and $N$ electrons in a
linearly polarized laser field. We suppose that the duration of the
laser pulse is much longer than the collision time and that the field
is homogeneous over the interaction region. We therefore adopt the
dipole approximation with a field described by the vector potential
\begin{equation}
\label{vecpot}
\Bm{A}(t)=\hat{\Bm z}\, A_0\cos\omega t
\end{equation}
where $\omega$ is the angular frequency and we have chosen the
$Z$-axis parallel to the direction of polarization.

The wave function for this system satisfies the Schr\"odinger equation
\begin{equation}
\label{SE}
\fl{\rm i} \frac{\partial}{\partial t} \Psi (\Bm{X}_{N+1},t) =
\left[ H_{N+1} - \frac{\zi}{\rmc} \Bm{A} (t) \cdot \sum_{e=1}^{N+1} 
\Bm{\nabla}_e + \frac{N+1}{2\rmc^2} \Bm{A}^2 (t) \right] \Psi 
(\Bm{X}_{N+1},t)
\end{equation}
where $\Bm{X}_{N+1}=\{\Bm{x}_1,\Bm{x}_2,\dots,{\Bm{x}}_{N},
{\Bm{x}}_{N+1} \}$ denotes the set of space- and spin-coordinates
of the $N+1$ electrons with $\Bm{x}_e=\{\Bm{r}_e,\Bm{\sigma}_e \}$, 
and $H_{N+1}$ is the field-free Hamiltonian operator
\[H_{N+1}=\sum_{e=1}^{N+1}\left[-\frac{1}{2}\Bm{\nabla}_e^2-\frac{Z}
{r_e}\right]+\sum_{e'>e=1}^{N+1}\frac{1}{|\Bm{r}_e-\Bm{r}_{e'}|}.\]

The presence of the linearly polarized laser field breaks the
spherical symmetry of the system by introducing a preferred direction
in space so that the total atomic angular momentum $L$ is no longer
defined. Its projection $M$ along the polarization axis is however
conserved since the system is invariant under rotation around this
axis. The dipole operator connects atomic states of opposite parities,
so that equation (\ref{SE}) separates for well-defined values of the
total atom-laser parity $\Pi$. As we neglect relativistic corrections,
the total atomic spin $S$ and projection $M_S$ are also good quantum
numbers. In what follows, we shall work in subspaces with fixed
quantum numbers $M\Pi SM_S$, which will usually not be explicitly
indicated.

As the Schr\"odinger equation (\ref{SE}) includes a time-dependent
potential which is periodic, it admits solutions in the form of a
Floquet-Fourier expansion
\begin{equation}
\label{FFexp}
\Psi(\XNp,t)=\e^{-\zi Et}\sum_{n=-\infty}^{\infty}\e^{-\zi n\omega t}
\Psi_n(\XNp)
\end{equation}
where $E$ is the quasi-energy of the solution. The conservation of
total parity $\Pi$ implies that each Floquet component in the
expansion (\ref{FFexp}) includes atomic wave functions of well-defined
parity, alternating with the parity of the Floquet index $n$. By
substituting (\ref{FFexp}) into (\ref{SE}) and projecting on a
particular Floquet component, we obtain an infinite set of
time-independent coupled equations for the $\Psi_n(\XNp)$ that can be
solved using standard methods.

Another general property of the Floquet-Fourier type solutions
(\ref{FFexp}) is the Shirley symmetry (Shirley 1965):
\begin{eqnarray}
\label{Shirley}
\fl
\Psi^E&=&\e^{-\zi Et}\sum_{n=-\infty}^{\infty} \e^{-\zi n\omega t}\Psi_n^E=
\e^{-\zi (E+m\omega)t}\sum_{n=-\infty}^{\infty}\e^{-\zi(n-m)\omega t}
\Psi_n^E\nonumber\\
\fl &=&\e^{-\zi (E+m\omega)t}\sum_{n=-\infty}^{\infty}\e^{-\zi n\omega t}
\Psi_{n+m}^E=\e^{-\zi (E+m\omega)t}\sum_{n=-\infty}^{\infty}\e^{-\zi n
\omega t}\Psi_{n}^{E+m\omega}.
\end{eqnarray}
This implies that equivalent sets of solutions of the Schr\"odinger
equation (\ref{SE}) exist, each characterised by quasi-energies with
period $\omega$ and Floquet components identical apart from a shift in
their indices.

\section{\textit{R}-matrix inner region}

In $R$-matrix theory, the inner region is defined as the portion of
configuration space where the radial coordinate of each electron is
smaller than the radius $a$ of the sphere encompassing the target
states. The treatment of the problem in this region, which is
analogous to that of atomic bound states, has been fully presented in
paper I. It is however useful to recall the equations that are
essential for understanding the treatment of laser-assisted scattering
in the outer region.

Since all the radial coordinates are small, it is appropriate to
define the wave function in the length gauge by the unitary transformation
\begin{equation}
\Psi (\XNp,t) = \exp \left( -\frac{\zi}{\rmc} \Bm{A}(t) \cdot \sum_{e=1}^{N+1} 
\Bm{r_e} \right) \Psi^{\rmL} (\XNp,t).
\end{equation}
$\Psi^{\rmL} (\XNp,t)$ is solution of the Schr\"odinger equation
\begin{equation}
\label{SEL}
\fl\zi\frac{\partial}{\partial t}\Psi^{\rmL}(\XNp,t)=\left[H_{N+1}
+\Bm{\cal E}(t)\cdot\sum_{e=1}^{N+1}\Bm{r_e}\right]\Psi^{\rmL}(\XNp,t)
\equiv H_{\mathrm F}^{\rmL}\Psi^{\rmL}(\XNp,t) 
\end{equation}
where the electric field $\Bm{\cal E}$ is defined as
\begin{equation}
\label{efield}
\Bm{\cal E}=-\frac{1}{\rmc}\frac{\rmd\Bm{A}}{\rmd t}=\hat{\Bm z}\, 
{\cal E}_0\,\sin\omega t
\end{equation}
with ${\cal E}_0=\omega A_0/\rmc$. By substituting the Floquet-Fourier 
expansion of $\Psi^{\rmL}(\XNp,t)$
\begin{equation}
\label{solL}
\Psi^{\rmL}(\XNp,t)=\e^{-\zi E^{\rmL} t}\sum_{n=-\infty}^{\infty}
\e^{-\zi n\omega t}\Psi_{n}^{\rmL}(\XNp)
\end{equation}
into equation (\ref{SEL}), multiplying by a particular $\exp(\zi
n\omega t)$ and integrating in time over a period $2\pi/\omega$, we
find that the Floquet components $\Psi_{n}^{\rmL}(\XNp)$ are solutions
of the coupled differential equations
\[\fl (H_{N+1}-E^{\rmL}-n\omega)\Psi_n^{\rmL}(\XNp)+D_{N+1}\Psi_{n+1}^{\rmL}
(\XNp)-D_{N+1}\Psi_{n-1}^{\rmL}(\XNp)=0\] 
with 
\[
D_{N+1}=\frac{{\cal E}_0}{2\zi}\,\hat{\Bm{z}}\cdot\sum_{e=1}^{N+1}\Bm{r_e}.
\]

We first determine a discrete $R$-matrix basis $\{\Psi_k^{\rmL}
(\XNp,t)\}$ satisfying the usual $R$-matrix boundary conditions. This
is done by diagonalizing the hermitian Hamiltonian $H_{\mathrm
F}^{\rmL}+ L_{\mathrm B}$, where $L_{\mathrm B}$ is a Bloch operator
cancelling the surface terms at $r=a$ arising from the kinetic
operators. The Floquet-Fourier components $\Psi_{kn}^{\rmL}(\XNp)$ of
the $R$-matrix basis functions are expanded in a set of fully
antisymmetrised $(N+1)$-electron functions which are formed by
coupling channel functions with defined symmetry $\Gamma=LM\pi SM_S$,
(where $\pi$ is the parity of the (N+1)-electron system)
\begin{eqnarray}
\label{chanf}
\bar{\Phi}_{i\ell}^{\Gamma}(\setminus r_{N+1})&=&\sum_{M_im}\sum_{M_{S_i}
\mu}\langle L_iM_i\ell m|LM\rangle\langle S_iM_{S_i}\frac{1}{2}\mu| SM_S
\rangle\nonumber\\
&&\times\Phi_i(\XN)Y_{\lm}(\theta_{N+1},\varphi_{N+1})\chi_{\frac{1}{2}\mu}
(\Bm{\sigma}_{N+1})
\end{eqnarray}
to continuum orbitals $u_{n\ell}(r_{N+1})$ which are chosen as regular
eigenfunctions of a field-free Hamiltonian including a static
potential of the target ground state, with zero log-derivative at
$r_{N+1}=a$. These orbitals are furthermore built to be orthogonal to those
in the target states $\Phi_i(\XN)$. $(N+1)$-electron bound
configurations constructed using the target orbitals are also included
in order to account for short-range correlation.

By projecting (\ref{solL}) onto the channel functions (\ref{chanf})
and onto the $n^{\mathrm th}$ component in photon space, and evaluating 
at the boundary $r_{N+1}=a$, we obtain the relation
\[F_{\nu}=\sum_{\nu'}R_{\nu \nu'} \; {\dot{F}}_{\nu'}\]
(where $\nu$ represents the set of indices $\{\Gamma i \ell n\})$ 
between the amplitudes of the radial functions
\[F_{\nu}=\langle\bar{\Phi}_{i\ell}^{\Gamma}|\delta(r_{N+1}-r_{\rm a})
\Psi_n^{\rmL}\rangle\]
and their derivatives
\[\dot{F}_{\nu}=\langle\bar{\Phi}_{i\ell}^{\Gamma}|\delta(r_{N+1}-r_{\rm a})
\frac{\partial}{\partial r_{N+1}}\Psi_n^{\rmL}\rangle\]
in terms of the $R$-matrix elements
\begin{equation}
\label{RMin}
R_{\nu\nu'}=\frac{1}{2}\sum_k\frac{\gamma_{\nu k}\gamma_{\nu'k}}{E_k-E}
\end{equation}
which depend on the quasi-energies $E_k$ and the surface amplitudes of
the eigenstates
\begin{equation}
\label{surfamp}
\gamma_{\nu k}=\langle\bar{\Phi}_{i\ell}^{\Gamma}|
\delta(r_{N+1}-r_{\rm a})\Psi_{kn}^{\rmL}\rangle.
\end{equation}
In principle the summation in (\ref{RMin}) is infinite, in practice it
is truncated. An approximation for the truncated part is provided by a
Buttle correction (Buttle 1967).

\section{Close-coupling equations using field-dressed target states}

The outer region is defined as the portion of configuration space where
the $N$ electrons of the target have a radial coordinate smaller than
$a$ while that of the collisional electron is larger. Exchange between
the collisional electron and the target electrons is negligible. From
now on, we can drop the index $N+1$ from the coordinates of the
collisional electron without possibility of confusion.

To solve the Schr\"odinger equation in the outer region, we adopt a
close-coupling expansion of the total wave function. Using the same
channel functions (\ref{chanf}) as in the inner region, we could write
\begin{equation}
\label{CCexp}
\Psi^{\rmL}(\XNp,t)=\e^{-\zi E^{\rmL}t}\sum_{n=-\infty}^{\infty}
\e^{-\zi n\omega t}\sum_{L\pi i\ell}\bar{\Phi}_{i\ell}^{\Gamma}
(\setminus r)\frac{1}{r}F_{\Gamma i\ell n}^{\rmL}(r)
\end{equation}
where the notation $\setminus r$ denotes the coordinates $\XN$ of all
the target electrons together with the angular and spin coordinates
of the collisional electron.
Substituting (\ref{CCexp}) into (\ref{SEL}), we find that the radial
functions $F_{\Gamma i\ell n}^{\rmL}(r)$ satisfy the set of coupled ordinary
differential equations in matrix notation
\begin{equation}
\label{CC}
\left(\frac{\rmd^2}{\rmd r^2}-\frac{\Bm{\ell}(\Bm{\ell}+\Bm{1})}{r^2}+
\frac{2z}{r}+\Bm{\kappa}^2\right)\bss{F}=\bss{W}(r)\bss{F}
\end{equation}
where $z=Z-N$ is the residual charge of the target. All quantities
operating on $\bss{F}$ on the left-hand side of (\ref{CC}) are diagonal
matrices. The elements of $\Bm{\kappa}^2$ are the square of the
channel momenta. The matrix $\bss{W}(r)$ includes the usual long-range
multipole potentials $\Bm{\alpha}^{(\lambda)}/r^\lambda$ arising from
the interaction of the collisional electron with the target, together
with the couplings of all the electrons with the laser field. The
interaction of the target with the field does not depend on the
coordinate of the collisional electron and therefore gives rise to
non-diagonal constant couplings in equation (\ref{CC}). In paper I
(appendix A2.1.2), these terms were removed by diagonalizing the
constant coupling matrix, which amounts to defining channel functions
with the field-dressed target states coupled to the angular and spin
wave functions of the collisional electron. This approach has two
disavantages. Couplings in the outer region are first defined in the
channel basis (\ref{chanf}) where their expression is cumbersome, then
transformed into the field-dressed target basis. Secondly, target
states differing only by the sign of their magnetic quantum number
$M_i$ are degenerate so that the diagonalization procedure can lead to
any linear combination of the solutions with $\pm|M_i|$.

We now show that the equations in the outer region are greatly
simplified by first determining the field-dressed target states and
using these to construct the channel functions. The magnetic quantum
numbers of the target and therefore of the collisional electron are
defined unambigously and scattering cross sections can be obtained in
a straightforward way.

The field-dressed target wave functions $\Psi^T(\XN,t)$ satisfy the
time-dependent Schr\"odinger equation (\ref{SEL}), with $N+1$
replaced by $N$:
\begin{equation}
\label{TSE}
\zi\frac{\partial}{\partial t}\Psi^T(\XN,t)=\left[H_{N}
+\Bm{\cal E}(t)\cdot\sum_{e=1}^N\Bm{r_e}\right]\Psi^T(\XN,t).
\end{equation}
We adopt the Floquet-Fourier expansion
\begin{equation}
\label{TFFexp}
\Psi^T(\XN,t)=\e^{-\zi E_T t}\sum_{n=-\infty}^{\infty}
\e^{-\zi n\omega t}\Psi_n^T(\XN)
\end{equation}
and expand each $\Psi_n^T(\XN)$ over the eigenfunctions $\Phi_i(\XN)$
of $H_N$ used in (\ref{chanf}):
\begin{equation}
\label{Tnexp}
\Psi_n^T(\XN)=\sum_{i}\Phi_i(\XN)a_{in}^T.
\end{equation}
The expansion (\ref{Tnexp}) includes atomic states with different
$L_i$ but the same $M_i$, $\pi_i$, $S_i$ and $M_{S_i}$. Substituting
(\ref{TFFexp}) and (\ref{Tnexp}) in (\ref{TSE}) and projecting on a
particular $\exp(-\zi n\omega t)\Phi_i(\XN)$, we find that the
coefficients $a_{in}^T$ are solutions of the eigenvalue problem
\[(w_i-E^T-n\omega)a_{in}^T+\sum_{i'}\langle\Phi_i\mid D_N\mid
\Phi_{i'}\rangle(a_{i'n+1}^T-a_{i'n-1}^T)=0,\] 
where we have used $H_N\Phi_i(\XN)=w_i\Phi_i(\XN)$. As a consequence
of the Shirley symmetry (\ref{Shirley}), the quasi-energies $E_T$ can
be grouped in sequences of period $\omega$.  The presence of the
linearly polarized field breaks the degeneracy in the magnetic quantum
number $M_T$ but states with the same value of $|M_T|$ are still
degenerate.  Since the Hamiltonian (\ref{TSE}) is non-relativistic,
states differing only by $M_{S_T}$ are degenerate.

We define channel functions with well-defined quantum numbers $M$,
$\Pi$, $S$ and $M_S$ by coupling the $\Psi^T(\XN,t)$ to angular and
spin functions of the collisional electron:
\begin{eqnarray}
\label{FDTchanf}
\fl\bar{\Phi}_{T\lm}(\setminus r,t)&=&\e^{-\zi E_Tt}\,\sum_{n=-\infty}
^{\infty}\,\e^{-\zi n\omega t}\,\sum_{M_{S_T}\mu}\langle S_TM_{S_T}
\frac{1}{2}\mu\mid SM_S\rangle\Psi_n^T(\XN)Y_{\lm}(\theta,\varphi)
\chi_{\frac{1}{2}\mu}(\Bm{\sigma}).\nonumber\\
\fl &&
\end{eqnarray}
In this basis, the close-coupling expansion of the wave function
$\Psi^{\rmL}(\XNp,t)$ is
\begin{equation}
\label{FDTCCexp}
\Psi^{\rmL}(\XNp,t)=e^{-\zi E^{\rmL}t}\sum_T\e^{\zi E_Tt}\sum_{\lm}
\bar{\Phi}_{T\lm}(\setminus r,t)\frac{1}{r}F_{T\lm}^{\rmL}(r).
\end{equation}
By comparing equations (\ref{CCexp}) and (\ref{FDTCCexp}), we find
that the radial functions $F_{T\lm}^{\rmL}(r)$ are related to those in
(\ref{CCexp}) by the transformation
\[F_{\Gamma i\ell'n}^{\rmL}(r)=\sum_{T\ell}O_{\Gamma i\ell'n,T\lm}
F_{T\lm}^{\rmL}(r)\]
with
\[O_{\Gamma i\ell'n,T\lm}=\langle L_i M_i\lm|LM\rangle a_{in}^{T}
\delta_{\ell'\ell}\,\delta_{M_i M_T}.\]
The Kronecker symbol $\delta_{\ell'\ell}$ arises since this basis
transformation concerns only the target and not the collisional
electron, while the second reflects the fact that the magnetic quantum
number of the target is not affected.
If we denote by ${\bss{F}}$ the vector of $F_{\Gamma i\ell n}^{\rmL}$
and by $\widetilde{\bss{F}}$ the vector of $F_{T\lm}^{\rmL}$, we have
in matrix notation
\[\bss{F}=\bss{O}\widetilde{\bss{F}}\] 
and similarly for their derivatives
\[\dot{\bss{F}}=\bss{O}\dot{\widetilde{\bss{F}}}.\] 
Since the $R$-matrix (\ref{RMin}) corresponding to the channels (\ref{chanf})
relates $\bss{F}$ and $\dot{\bss{F}}$ by
\[\bss{F}=\bss{R}\dot{\bss{F}},\]
the $R$-matrix $\widetilde{\bss{R}}$ corresponding to the field-dressed
channels (\ref{FDTchanf}) is related to $\bss{R}$ by the transformation
\[\widetilde{\bss{R}}=\bss{O}^{\mathrm t}\bss{R}\bss{O}\] 
where $\bss{O}^{\mathrm t}$ is the transpose of $\bss{O}$.  The
long-range multipole potential coefficients in equation (\ref{CC})
must also be transformed in a similar way:
\[\widetilde{\Bm{\alpha}}^{(\lambda)}=\bss{O}^{\mathrm t}\Bm{\alpha}
^{(\lambda)}\bss{O}.\]
The channel energies $\Bm{\kappa}^2/2$ must be recalculated in this new
basis, while the centrifugal term remains the same.

\section{Length to velocity gauge transformation and propagation in the 
velocity gauge} 
In the outer region, the potential in the length gauge $\Bm{\cal E}(t)
\cdot\Bm{r}$ diverges as the radial coordinate of the collisional
electron increases. It is therefore more appropriate to express the
interaction of this electron with the laser field in the velocity
gauge, where the potential remains finite. We conveniently perform
this gauge transformation at $r=a$, although in principle it could be
done at a larger distance. The relation between the wave function in
the length gauge $\Psi^{\rmL}(\XNp,t)$ and that in the velocity gauge
$\Psi^{\rmV}(\XNp,t)$ is
\begin{equation}
\label{wfV}
\fl\Psi^{\rmV}(\XNp,t)=\exp\left\{\frac{\zi}{2c^2}\int^t
A^2(t'){\mathrm d}t'-\frac{\zi}{c}\Bm{A}(t)\cdot\Bm{r}\right\}
\Psi^{\rmL}(\XNp,t).
\end{equation}
The Schr\"odinger equation in the velocity gauge is
\begin{eqnarray}
\label{SEV}
\fl\zi\frac{\partial}{\partial t}\Psi^{\rmV}(\XNp,t)&=&\left[H_{N}
+\Bm{\cal E}(t)\cdot\sum_{e=1}^N\Bm{r_e}\right.\nonumber\\
&&\left.-\frac{1}{2}\nabla^2-\frac{Z}{r}+\sum_{e=1}^N\frac{1}
{\mid\Bm{r}-\Bm{r_e}\mid}
-\frac{\zi}{c}\Bm{A}(t)\cdot\Bm{\nabla} \right]\Psi^V(\XNp,t).
\end{eqnarray}
Analogous to (\ref{FDTCCexp}), we expand $\Psi^{\rmV}(\XNp,t)$ as
\begin{equation}
\label{CCexpV}
\Psi^{\rmV}(\XNp,t)=\e^{-\zi E^{\rmV}t}\sum_T\e^{\zi E_Tt}
\sum_{\lm}\bar{\Phi}_{T\lm}(\setminus r,t)\frac{1}{r}F_{T\lm}^{\rmV}(r).
\end{equation}
Substituting (\ref{CCexpV}) into equation (\ref{SEV}) and projecting
on a particular channel function (\ref{FDTchanf}), we obtain the
close-coupling equations in the velocity gauge. In matrix form, they
are
\begin{equation}
\label{CCV}
\left(\frac{\rmd^2}{\rmd r^2}-\frac{\Bm{\ell}(\Bm{\ell}+\Bm{1})}{r^2}+
\frac{2z}{r}+\Bm{\kappa}^2+\bss{P}\frac{\rmd}{\rmd r}\right)\bss{F}^{\rmV}=
\sum_{\lambda=1}^{\lambda_{\mathrm{max}}}\frac{\Bm{\widetilde{\alpha}}
^{(\lambda)}}{r^{\lambda}}\bss{F}^{\rmV}
\end{equation}
where $\bss{P}\rmd/\rmd r$ and $\Bm{\widetilde{\alpha}}^{(1)}/r$
arise from the term $\Bm{A}(t)\cdot\Bm{\nabla}$. The elements of
$\bss{P}$ are
\begin{eqnarray*}
\fl P_{T\lm,T'\ell'm'}&=&\frac{A_0}{c}(\delta_{T,T'-\omega}+
\delta_{T,T'+\omega})\delta_{mm'}\\
\fl&&\times\left\{-\sqrt{\frac{(\ell'-m)(\ell'+m)}{(2\ell'+1)(2\ell'-1)}}
\delta_{\ell,\ell'-1}+\sqrt{\frac{(\ell-m)(\ell+m)}
{(2\ell+1)(2\ell-1)}}\delta_{\ell,\ell'+1}\right\}
\end{eqnarray*}
while those of $\Bm{\widetilde{\alpha}}^{(1)}$ are
\begin{eqnarray*}
\fl \Bm{\widetilde{\alpha}}^{(1)}_{T\lm,T'\ell'm'}&=&\frac{A_0}{c}
(\delta_{T,T'-\omega}+\delta_{T,T'+\omega})\delta_{mm'}\\
\fl&&\times\left\{\ell'\sqrt{\frac{(\ell'-m)(\ell'+m)}{(2\ell'+1)(2\ell'-1)}}
\delta_{\ell,\ell'-1}+\ell\sqrt{\frac{(\ell-m)(\ell+m)}
{(2\ell+1)(2\ell-1)}}\delta_{\ell,\ell'+1}\right\}.
\end{eqnarray*}
The Kronecker symbol $\delta_{T,T'\pm\omega}$ means that $T$ and $T'$
are the same field-dressed target states ({\sl i.e.} with the same
atomic structure), differing only by their quasi-energies
$E_T=E_{T'}\pm\omega$. The dipole operator couples channels where the
collisional electron absorbs or emits one photon. $\bss{P}$ and
$\Bm{\widetilde{\alpha}}^{(1)}$ obey the usual one-electron dipole
selection rules $\Delta\ell =\pm1$ and $\Delta m=0$. The fact that
$\bss{P}$ and $\Bm{\widetilde{\alpha}}^{(1)}$ are real is a consequence of
the choice of phase for the field and the Fano-Racah phase convention
adopted for the spherical harmonics. Since the Hamiltonian is
hermitian, $\bss{P}$ is antisymmetric and $\Bm{\widetilde{\alpha}}^{(1)}$
is symmetric.

In order to solve the close-coupling equations (\ref{CCV}) we must
first transform $\widetilde{\bss{R}}$ into the velocity gauge. Using
equations (\ref{vecpot}) and (\ref{wfV}), we write
\begin{eqnarray}
\label{L2V}
\fl\Psi^{\rmV}(\XNp,t)&=&\exp\left\{\zi\frac{A_0^2}{4c^2}t\right\}
\exp\left\{\zi\frac{A_0^2}{8\omega c^2}\sin2\omega t-\frac{\zi}{c}A_0
\hat{\Bm{z}}\cdot\Bm{r}\cos\omega t\right\}\Psi^{\rmL}(\XNp,t).\nonumber\\
\fl
\end{eqnarray}
By comparing this with equation (\ref{CCexpV}) and using (\ref{solL}), 
we may first identify
\[E^{\rmV}=E^{\rmL}-\frac{A_0^2}{4c^2}=E^{\rmL}-E_{\mathrm P}\]
where $E_{\mathrm P}$ is the ponderomotive energy. The second
exponential in equation (\ref{L2V}) is periodic in time and can be 
written as the product of two Fourier series,
\begin{eqnarray}
\label{Fexp}
\fl\exp\left\{\zi\frac{A_0^2}{8\omega c^2}\sin2\omega t-\frac{\zi}{c}A_0
\hat{\Bm{z}}\cdot\Bm{r}\cos\omega t\right\}&=&
\sum_{{\cal N}}\,\e^{-\zi2{\cal N}\omega t}J_{-{\cal N}}
\left(\frac{A_0^2}{8\omega c^2}\right)\nonumber\\
&&\times\sum_{{\cal N}'}\,\e^{-\zi{\cal N}'\omega t}\,\zi^{\cal{N}'}
J_{\cal{N}'}\left(-\frac{A_0}{c}\hat{\Bm{z}}\cdot\Bm{r}\right).
\end{eqnarray}
Substituting equation (\ref{Fexp}) into (\ref{L2V}), replacing
$\Psi^{\rmV}$ and $\Psi^{\rmL}$ by their respective close-coupling
expansions (\ref{FDTCCexp}) and (\ref{CCexpV}) and projecting onto the
channel functions (\ref{FDTchanf}), we obtain relations between the
radial functions $F_{T\lm}^{\rmV}(r)$ and $F_{T\lm}^{\rmL}(r)$. These
can be written in matrix form as
\begin{equation}
\label{FV}
\bss{F}^{\rmV}=\bss{A}\widetilde{\bss{F}}
\end{equation}
where the elements of $\bss{A}$ are given by
\begin{eqnarray}
\label{L2Vmat}
\fl A_{T\lm,T'\ell'm'}&=&\sum_{{\cal N}}\,\zi^{\ell'-\ell-k-2{\cal N}}
J_{-{\cal N}}\,\left(\frac{A_0^2}{8\omega c^2}\right)\,
\left[\frac{(2\ell+1)(2\ell'+1)(\ell-m)!(\ell'-m')!}
{(\ell+m)!(\ell'+m')!}\right]^{\frac{1}{2}}\nonumber\\
\fl&&\times\int_0^{\pi/2}{\mathrm d}\theta\sin\theta\,P_{\ell}^{m}
(\cos\theta)P_{\ell'}^{m'}(\cos\theta)
J_{-k-2{\cal N}}\left(-\frac{A_0}{c}r\cos\theta\right)\,
\delta_{mm'}.\nonumber\\
\fl
\end{eqnarray}
Since the gauge transformation applies only to the scattered electron,
the channels coupled by the matrix $\bss{A}$ are those where the
atomic structure of the field-dressed target state is the same:
$E_T=E_{T'}+k\omega$, where $k$ is an integer. The Kronecker symbol
$\delta_{mm'}$ arises from the fact that the laser field is polarized
along the $Z$-axis.  The integral in equation (\ref{L2Vmat}) is
non-zero only if $\ell+\ell'-k-2{\cal N}$ is even, implying that the
matrix $\bss{A}$ is real.

Since the radial functions $\bss{F}^{\rmV}$ and $\widetilde{\bss{F}}$ are
related by the simple linear transformation (\ref{FV}), it is easy to
show that
\[\bss{F}^{\rmV}\left(\dot{\bss{F}}^{\rmV}\right)^{-1}=
\bss{A}\left(\dot{\bss{A}}+\bss{A}\widetilde{\bss{R}}^{-1}\right)^{-1}\]
where the elements of $\dot{\bss{A}}$ are the derivatives of
(\ref{L2Vmat}) at $r=a$. In order to apply standard propagation
techniques to solve (\ref{CCV}), we remove the $\bss{P}\rmd/\rmd r$ 
terms by transforming the radial functions as
\begin{equation}
\label{Smith}
\bss{F}^{\rmV}=\e^{-\bss{P}r/2}\bss{G}.
\end{equation}
This is analogous to a Smith-type diabatization (Smith 1969) in
molecular scattering. Substituting (\ref{Smith}) into (\ref{CCV}), we
find that the functions $\bss{G}$ satisfy the set of coupled equations
\[\fl
\left(\frac{\rmd^2}{\rmd r^2}-\frac{\bss{P}^2}{4}+
\e^{\bss{P}r/2}\left[-\frac{\Bm{\ell}(\Bm{\ell}+\Bm{1})}{r^2}+
\frac{2z}{r}+\Bm{\kappa}^2-\sum_{\lambda=1}^{\lambda_{\mathrm{max}}}
\frac{\Bm{\widetilde{\alpha}}^{(\lambda)}}{r^{\lambda}}\right]
\e^{-\bss{P}r/2}\right)\bss{G}=0
\]
where the couplings oscillate with the distance $r$. The $R$-matrix
corresponding to the functions $\bss{G}$ is given by
\[\bss{R}_G=\e^{\bss{P}r/2}\left(\dot{\bss{F}}^{\rmV}\left(\bss{F}^{\rmV}
\right)^{-1}+\frac{\bss{P}}{2}\right)^{-1}\e^{-\bss{P}r/2}\]
which is symmetric. A detailed discussion of the properties of
$\bss{R}_G$ has been presented in paper II.

To propagate the matrix $\bss{R}_G$, we first inverse it to obtain the
log-derivative matrix. We propagate this outwards using the method of
Johnson and Manolopoulos (Manolopoulos 1986) which allows oscillating
potentials to be treated with relatively large propagation steps.
The propagated log-derivative matrix is then inverted back to the 
matrix $\bss{R}_G$.

\section{Scattering boundary conditions and asymptotic solutions}

After propagating the $R$-matrix in the velocity gauge, we need to
match to the asymptotic solutions corresponding to the scattering
boundary conditions. These can easily be defined if the close-coupling
equations are asymptotically uncoupled, which is not the case in the
velocity gauge. We apply the Kramers-Henneberger transformation
(Kramers 1956, Henneberger 1968)
\begin{equation}
\label{V2A}
\fl\Psi^{\rmV}(\XNp,t)=\exp[-\zi\Bm{\alpha}(t)\cdot\Bm{p}]\,
\Psi^{\mathrm A}(\XNp,t)\equiv\Psi^{\rmA}(\XN,\Bm{r}+\Bm{\alpha}(t),
\Bm{\sigma},t)
\end{equation}
where 
\[\Bm{\alpha}(t)\,=\,\frac{1}{c}\int^t\Bm{A}(t')\rmd t'\,=\,\hat{\Bm{z}}\,
\alpha_0\sin\omega t=\hat{\Bm{z}}\,{\cal E}_0/\omega^2 \sin\omega t\] 
is the oscillation vector of a free electron in the laser field. It
must be emphasized that this transformation is applied only to the
collisional electron. The target electrons, which are tightly bound to
the nucleus, are kept in the length gauge. By substituting (\ref{V2A})
into (\ref{SEV}), we obtain the close-coupling equations in
the acceleration frame:
\begin{eqnarray}
\label{SEA}
\fl\zi\frac{\partial}{\partial t}\Psi^{\mathrm A}(\XNp,t)&=&\left[H_{N}
+\Bm{\cal E}\cdot\sum_{e=1}^N\Bm{r}_e\right.\nonumber\\
&&\left.-\frac{1}{2}\nabla^2-\frac{Z}{\mid\Bm{r}+\Bm{\alpha}(t)\mid}+
\sum_{e=1}^N\frac{1}{\mid\Bm{r}+\Bm{\alpha}(t)-\Bm{r_e}\mid}\right]
\Psi^A(\XNp,t).
\end{eqnarray}
In the limit $r\rightarrow\infty$, $\alpha_0$ and $r_e$ are negligible
with respect to $r$. The right-hand side of (\ref{SEA}) reduces to the
target Floquet Hamiltonian plus the Hamiltonian of an electron in the
screened Coulomb potential of residual charge $z=Z-N$. Two sets of
linearly independent solutions can therefore be defined, with boundary
conditions
\begin{eqnarray*}
\fl\Psi_{\nu}^{\rmA}(\XNp,t) &\rlim&\e^{-\zi E^{\rmA}t}\sum_{M_{S_T}}
\sum_{\mu} (S_TM_{S_T}\frac{1}{2}\mu|SM_S)\e^{\zi E_Tt}\Psi^{T}(\Bm{X}_N,t)\\ 
\fl &&\times\frac{1}{\sqrt{k_{\nu}}}\frac{1}{r}\,\e^{\pm\zi\theta_{\nu}(r)}
Y_{\ell m}(\theta,\varphi)\;\chi_{\frac{1}{2}\mu}(\Bm{\sigma})
\end{eqnarray*}
which can be rewritten as
\begin{eqnarray}
\label{BCA}
\fl\Psi_{\nu}^{\rmA}(\XNp,t) &\rlim&\sum_{M_{S_T}}\sum_{\mu}
(S_TM_{S_T}\frac{1}{2}\mu|SM_S)\Psi^{T}(\Bm{X}_N,t)
\nonumber\\
\fl &&\times\e^{-\zi\epsilon_{\nu}t}\frac{1}{\sqrt{k_{\nu}}}
\frac{1}{r}\,\e^{\pm\zi\theta_{\nu}(r)}Y_{\ell m}(\theta,\varphi)\;
\chi_{\frac{1}{2}\mu}(\Bm{\sigma})
\end{eqnarray}
where $\epsilon_{\nu}=E^{\rmA}-E_{T}=k_{\nu}^2/2$ is the channel
energy of the scattered electron in the acceleration frame.  From
equation (\ref{V2A}) and the Floquet expansions of
$\Psi^{\rmA}(\XNp,t)$ and $\Psi^{\rmV}(\XNp,t)$, it is obvious that
the quasi-energy in the acceleration frame $E^{\rmA}$ is equal to
$E^{\rmV}$. We have introduced the collective index $\nu=\{T,\ell,m\}$
to identify the channel. The Coulomb phase is
$\theta_{\nu}(r)=k_{\nu}r-\ell\pi/2-\eta_\nu\ln(2k_{\nu}r)+\sigma_{\ell}
(\eta_\nu)$, with $\eta_{\nu}=-z/k_{\nu}$ and $\sigma_{\ell}
(\eta_{\nu})=\arg\Gamma(\ell+1+\zi\eta_{\nu})$.

We now define the boundary conditions in the velocity gauge by
applying the frame transformation (\ref{V2A}) to (\ref{BCA}):
\begin{eqnarray} 
\label{BCV}
\fl\Psi_{\nu}^{\rmV}(\XNp,t) &\rlim&\sum_{M_{S_T}}\sum_{\mu}
(S_TM_{S_T}\frac{1}{2}\mu|SM_S)\Psi^{T}(\Bm{X}_N,t)\nonumber\\ 
\fl &&\times\e^{-\zi\epsilon_{\nu}t}\frac{1}{\sqrt{k_{\nu}}}
\frac{1}{r_{\alpha}}\,\e^{\pm\zi\theta_{\nu}(r_{\alpha})}Y_{\ell m}
(\theta_{\alpha},\varphi)\;\chi_{\frac{1}{2}\mu}(\Bm{\sigma})
\end{eqnarray}
where $(r_{\alpha},\theta_{\alpha},\varphi)$ are the spherical
coordinates of the shifted vector $\Bm{r}_{\alpha}(t)=\Bm{r}+
\Bm{\alpha}(t)$. In the limit $r\rightarrow\infty$, we have
\[r_{\alpha}=r-\alpha_0\sin\omega t\cos\theta+{\cal O}\left(\frac{1}
{r}\right)\]
so that
\[\frac{1}{r_{\alpha}}=\frac{1}{r}+{\cal O}\left(\frac{1}{r^2}\right).\]
The azimuthal angle of $\Bm{r}_{\alpha}$ is simply given by
\[\cos\theta_{\alpha}=\cos\theta+{\cal O}\left(\frac{1}{r}\right).\]
We can therefore use the approximation
\begin{eqnarray*}
\fl\exp \{\zi k r_\alpha - \zi \eta \ln(2 k r_\alpha)\}
=\exp \{ \zi k r - \zi \eta \ln(2 k r) \} \; 
\exp \{ -\zi k \alpha_0 \sin \omega t \; \cos \theta \} 
\; + {\cal O}\left(\frac{1}{r}\right).
\end{eqnarray*}
The time-dependence in the phase $\theta_\nu(r_{\alpha})$ leads to a
mixing of the channels in the velocity gauge. We may write
equation (\ref{BCV}) as
\begin{equation} 
\label{BCV2}
\Psi_{\nu}^{\rmV}(\XNp,t)\rlim 
\sum_{\nu'}\bar{\Phi}_{\nu'}(\setminus r,t)\rmA_{\nu'\nu}^{(0)}
\e^{-\zi\epsilon_{\nu}t}\frac{1}{\sqrt{k_{\nu}}}\frac{1}{r}\,
\e^{\pm\zi\theta_{\nu}(r)}
\end{equation}
where $\nu'=\{T',\ell',m'\}$.
By equating (\ref{BCV}) and (\ref{BCV2}) and using
\[
J_n(x)=\frac{1}{2\pi}\int_0^{2\pi}\rmd\theta\;\e^{\zi(x\sin\theta-n\theta)}
\]
we find that
\[\rmA_{\nu'\nu}^{(0)}=\int\rmd\Omega\;Y_{\ell'm'}^{\ast}
(\theta,\varphi)\; Y_{\ell m}(\theta,\varphi)\;J_k(\mp k_{\nu}\alpha_0
\cos\theta)\;\delta_{m'm}\]
where $k$, defined by $E_{T'}=E_{T}+k\omega$, must be an
integer if $\rmA_{\nu'\nu}^{(0)}$ is to be non-zero. This is again a
consequence of the fact that the frame transformation is applied only
to the scattered electron. The magnetic quantum numbers are conserved
since the transformation is along the Z-direction.

We now use (\ref{BCV2}) as the first term of an asymptotic expansion
of the radial functions in the velocity gauge
\begin{equation}
\label{AEV}
\Psi_{\nu}^{\rmV}(\XNp,t)=\sum_{\nu'}\bar{\Phi}_{\nu'}(\setminus r,t)
\sum_{\mu=0}^{\mu_{\mathrm{max}}}\frac{1}{r^{\mu}}A_{\nu'\nu}^{(\mu)}\,
\e^{-\zi\epsilon_{\nu}t}\frac{1}{\sqrt{k_{\nu}}}\frac{1}{r}\,
\e^{\pm\zi\theta_{\nu}(r)}.
\end{equation}
In matrix notation, we have for the radial functions
\begin{equation}
\label{AEVM}
\bss{F}^{\rmV}(r)=\frac{1}{\sqrt{\bss{k}}}\frac{1}{r}\,
e\,^{\pm\zi\,{\btheta}(r)}\;
\sum_{\mu=0}^{\mu_{\mathrm{max}}}  \frac{1}{r^{\mu}}\,\bss{A}^{(\mu)}
\end{equation}
where the factor in front of the summation denotes a diagonal matrix
with elements $\exp(\pm\zi\theta_\nu(r))/\sqrt{k_\nu}r$. By substituting
(\ref{AEVM}) into (\ref{CCV}), we obtain a set of recurrence relations
for $\bss{A}^{(\mu)}$. Details have been given in paper II.

\section{Scattering amplitudes and cross sections}
The close-coupling equations in the acceleration frame are asymptotically
uncoupled, as in the field-free case. We can therefore impose the
usual scattering boundary conditions
\[
\begin{array}{llll}
F_{\nu\nu'}(r)&\!\!\rlim&\!\!k_{\nu}\sp{-1/2}(\sin\theta_{\nu}
\delta_{\nu\nu'} +\cos\theta_{\nu}K_{\nu\nu'}),&\quad k_{\nu}\sp2>0\\ 
F_{\nu\nu'}(r)&\!\!\rlim&\!\! \e^{-|k_{\nu}|r},&\quad k_{\nu}\sp2<0.
\end{array}
\]
The $K$-matrix is calculated by matching the solutions propagated from
the inner region to the asymptotic solutions in the velocity
gauge. The matching equation is obtained by writing the propagated
solutions as
\[G_{\nu\nu'}^{\rmV}(r)=\sum_{\nu''=1}^{n+\no}H_{\nu\nu''}(r)\;
x_{\nu''\nu'}\]
where the elements of $\bss{H}$ are the asymptotic solutions with
boundary conditions in the acceleration gauge
\[
\fl
\begin{array}{llll}
H_{\nu\nu'}(r)&\!\!\rlim&\!\!\sin\theta_{\nu}\,\delta_{\nu\nu'}&
\quad \nu=1,\dots n; \; \nu'=1,\dots \no\\ 
H_{\nu\nu'}(r)&\!\!\rlim&\!\!\cos\theta_{\nu}\,\delta_{\nu\nu'-\no}&
\quad \nu=1,\dots n; \; \nu'=\no+1,\dots 2\no\\
H_{\nu\nu'}(r)&\!\!\rlim&\!\!\e^{-|k_{\nu}|r}\,\delta_{\nu\nu'-\no}&
\quad \nu=1,\dots n; \; \nu'=2\no+1,\dots n+\no.
\end{array}
\]
The first two sets correspond respectively to the imaginary and real
parts of the solutions (\ref{AEV}) with $k_{\nu}^2>0$. For the open
channels, we need to calculate only one of the $\exp(\pm\zi k_{\nu}r)$
solutions. The third set corresponds to the solutions (\ref{AEV}) with
$k_{\nu}^2<0$. For the closed channels, $k_{\nu}$ is imaginary and we
calculate the solutions decaying asymptotically. The matrix $\bss{H}$
is of dimension $n\times(n+\no)$, where $n$ is the total number of
channels and $\no$ is the number of open channels. The $K$-matrix is
obtained from the solution $\bss{X}$ to the equation
\[\bss{M}_1\bss{X}=\bss{M}_2\]
where $\bss{M}_2$ and $\bss{M}_1$ correspond respectively to the
first $\no$ and the last $n$ columns of the matrix
\[\bss{H}- \bss{R}_G\,\dot{\bss{H}}-\bss{R}_G\frac{\bss{P}}{2}\bss{H}.\]
The matrix $\bss{X}$ has dimension $n\times\no$ and the $K$-matrix is
given by the first $\no$ rows of this matrix. 

The scattering amplitudes are obtained from the $K$-matrix in a manner
similar to the field-free case (see for example Smith 1971). 
Differences arise due to the presence of the laser field which
breaks the spherical symmetry. We first suppose that the target is
neutral and that the incoming electron has a momentum $\Bm{k}_i$ and a
spin projection $\mu_i$ while the outgoing electron has a
momentum $\Bm{k}_f$ and a spin projection $\mu_f$. The scattering 
amplitude for a transition from the initial field-dressed target 
state $i$ to the final field-dressed target state $f$ is found to be
\begin{eqnarray}
\fl f_{i\mu_i, f\mu_f}(\Bm{k}_i,\Bm{k}_f) & = &
\frac{2\pi\zi}{\sqrt{k_i k_f}} \sum_S\,\langle S_i M_{S_i} \frac{1}{2}
\mu_i \mid S M_S\rangle \langle S_f M_{S_f} \frac{1}{2} \mu_f \mid S
M_S\rangle\nonumber\\ 
\fl&&\times\sum_{M \Pi}\,\sum_{\ell\ell'}\,Y_{\ell m}^*(\hat{\Bm{k}}_i)
Y_{\ell'm'}(\hat{\Bm{k}}_f)\,T_{i\ell m, f\ell'm'}^{M\Pi S}.
\label{famp}
\end{eqnarray}
$T_{i\ell m, f\ell'm'}^{M\Pi S}$ is an element of the $T$-matrix which
is related to the $S$-matrix by $\bss{T} = \bss{1} - \bss{S}$, with
$\bss{S} = (\bss{1} + \zi\bss{K})(\bss{1} - \zi\bss{K})^{-1}$. The
$T$-matrix is block-diagonal with respect to the three good quantum
numbers $M$, $\Pi$ and $S$. It is independent of $M_S$ and the sign of
$M$. The expression (\ref{famp}) is simpler than that in the
field-free case since the total angular momentum $L$ is no longer a
good quantum number. It is therefore not necessary to uncouple the
angular momenta of the target and scattered electron to define the
scattering amplitude.

The differential cross section for the scattering of an unpolarized
electron by an unpolarized target is obtained by summing the square of
the scattering amplitudes over all possible final states and averaging
over all initial states:
\begin{eqnarray}
\fl
\frac{\rmd\sigma_{if}}{\rmd\Omega}&=&\frac{k_f}{k_i}\frac{1}{2(2S_i+1)}
\sum_{M_{S_i}\mu_iM_{S_f}\mu_f}\mid f_{i\mu_i,f\mu_f}
\mid^2\nonumber\\
\fl&=&\frac{\pi^2}{E_i}\frac{1}{2S_i+1}\sum_S (2S+1)
\left|\sum_{M\Pi}\sum_{\ell\ell'}\,Y_{\ell m}^*(\hat{\Bm{k}}_i)
Y_{\ell'm'}(\hat{\Bm{k}}_f)\,T_{i\ell m,f\ell'm'}^{M\Pi S}\right|^2
\label{difxs}
\end{eqnarray}
where $E_i$ is the kinetic energy of the incoming electron.
The total cross section is obtained by integrating (\ref{difxs}) over
the outgoing angles
\begin{equation}
\fl\sigma_{if}=\frac{\pi^2}{E_i}\frac{1}{2S_i+1}\sum_S (2S+1)
\sum_{M\Pi}\sum_{\ell\ell'\ell''}\,Y_{\ell m}^*(\hat{\Bm{k}}_i) 
Y_{\ell''m}(\hat{\Bm{k}}_i)\,T_{i\ell m,f\ell'm'}^{M\Pi S}
\,\left(T_{i\ell''m,f\ell'm'}^{M\Pi S}\right)^*
\end{equation}
where we have $m=M-M_i$ and $m'=M-M_f$.  

In the case of an ionic target with residual charge $z$, the
scattering amplitude contains two terms
\begin{equation}
\label{fampion}
 f_{i\mu_i, f\mu_f}(\Bm{k}_i,\Bm{k}_f)=
f_C(\Bm{k}_i,\Bm{k}_f)\,\delta_{if}+f_{i\mu_i,
f\mu_f}^{(\mathrm{rad})}(\Bm{k}_i,\Bm{k}_f).
\end{equation}
The first term is the Coulomb scattering amplitude
\[f_C(\Bm{k}_i,\Bm{k}_f)=\frac{z}{2k^2\sin^2(\Theta/2)}{\mathrm e}^
{2\zi\sigma_0(k)}{\mathrm e}^{\zi(z/k)\ln[\sin^2(\Theta/2)]}\]
where $\Theta$ is the angle between $\Bm{k}_i$ and $\Bm{k}_f$ and
$k$ is the amplitude of $\Bm{k}_i$ and $\Bm{k}_f$.
The second term is similar to (\ref{famp}) and has the form
\begin{eqnarray}
\fl f_{i\mu_i, f\mu_f}^{(\mathrm{rad})}(\Bm{k}_i,\Bm{k}_f)&=&
\frac{2\pi\zi}{\sqrt{k_i k_f}} \sum_S\,\langle S_i M_{S_i} \frac{1}{2}
\mu_i \mid S M_S\rangle \langle S_f M_{S_f} \frac{1}{2} \mu_f \mid S
M_S\rangle\nonumber\\ 
\fl&&\times\sum_{M \Pi}\,\sum_{\ell\ell'}\, {\mathrm
e}^{\zi[\sigma_{\ell}(k_i)+\sigma_{\ell'}(k_f)]} Y_{\ell
m}^*(\hat{\Bm{k}}_i) Y_{\ell'm'}(\hat{\Bm{k}}_f)\,T_{i\ell m,
f\ell'm'}^{M\Pi S}.
\label{famprad}
\end{eqnarray}
The differential cross section is obtained by taking the modulus
square of (\ref{fampion}), averaging over the initial states and
summing over the final states. As in the field-free case, the total
elastic cross section is not defined as it diverges.

It should be stressed that the above formulae are for transitions
between field-dressed target states. The Shirley symmetry implies that
these can be grouped in equivalence classes, each with the same atomic
structure and a quasi-energy spectrum of period $\omega$.  When the AC
Stark mixing is weak, these quasi-energies will lie close to the
field-free thresholds shifted by an integer multiple of the photon
energy. It is then natural, {\em though not necessary}, to use these
integers as labels, with the field-dressed states closest to the
field-free ones defining the ``zero-photon'' thresholds.  A transition
from state $(a,M)$ to state $(b,N)$ can therefore be interpreted as
the atomic transition $a\rightarrow b$ with net exchange of $|M-N|$
photons. The collisional electron emerges with a kinetic energy
differing by $|M-N|\omega$ from the kinetic energy associated with the
zero-photon transition.  When the AC Stark mixing is very strong
however, it may be difficult to identify the field-dressed threshold
as a field-free threshold shifted by a particular number of photon
energies. For inelastic processes, we may choose a particular
transition between two field-dressed states $i$ and $f$ as the
``zero-photon'' transition for the associated atomic states. This can
then be used as a reference for counting the number of photons
exchanged for transitions between other members of the two sequences
containing $i$ and $f$.

\section{Computational implementation}

An advantage of the $R$-matrix Floquet theory is that it can re-use
existing $R$-matrix inner region computer codes for field-free
electron-atom collisions and photoionization that have been developed
over many years (Berrington \etal 1978, 1987). The Fano-Racah phase
convention for the spherical harmonics used in those packages has to
be kept, which yields rather unusual expressions for the dipole matrix
elements. This also justifies the choice of phase for the laser field 
(\ref{efield}), since a real symmetric Floquet Hamiltonian is more
convenient from the computational point of view.

The first task is to define a set of field-free target wave functions.
While in principle any method could be used to determine these, in
practice only the structure programs {\tt CIV3} (Hibbert 1975) and
{\tt SuperStructure} (Eissner \etal 1974) are interfaced with the
$R$-matrix codes. Once the target state energies, wave functions and
dipole matrices have been obtained, the field-free $R$-matrix inner
region codes are used to calculate dipoles, energies and surface
amplitudes for the (N+1)-electron system.  The next step, {\tt MPB},
builds and diagonalizes the Floquet Hamiltonian matrix $H_{\mathrm
F}^{\rmL}+ L_{\mathrm B}$ and calculates the surface amplitudes
(\ref{surfamp}) required to construct the $R$-matrix (\ref{RMin}) in
the inner region. In the current implementation, this stage is
parallelized using {\tt SCALAPACK} (Blackford \etal 1997). The above
steps need only be performed once for fixed laser frequency and
intensity as they are independent of the collision energy.

The program {\tt MPSCAT} solves the scattering problem in the outer
region. For each collision energy, the $R$-matrix from the inner
region is transformed into the field-dressed channel representation,
and from the length into the velocity gauge. Its inverse is then
propagated from the inner region boundary out to a specified distance
using a log-derivative propagator (Manolopoulos 1986). The propagated
solutions are matched with those from the asymptotic expansion
(\ref{AEV}) to give the $K$-matrix, from which cross sections can be
calculated.

In the field-free case, the $R$-matrix is symmetric by construction.
This is also the case in the length gauge, but not in the velocity
gauge since the Floquet expansion is truncated and the transformation
matrix $\bss{A}$ (\ref{L2Vmat}) is not quite unitary. Adopting the
velocity gauge in the inner region would yield a symmetric $R$-matrix
but the Buttle correction then requires non-diagonal elements that are
difficult to evaluate.  The propagation method we employ however does
not suppose the matrix to be symmetric.  The transformation from the
acceleration frame into the velocity gauge is not unitary for the same
reason. As a consequence, the $K$-matrix is not symmetric. In
practice, we find that this does not affect the cross section,
provided the {\em full} not quite symmetric $K$-matrix is used to
determine the $S$-matrix and the contributions of the channels with
the largest angular momenta or largest (absorption and emission)
Floquet indices are not included.

From the computational point of view, calculations for transitions
between two continuum states are more demanding than for bound-free
transitions since more partial waves need to be included. It is also
necessary to integrate over large interaction distances, especially to
obtain converged angular distributions. Many operations can be written
in matrix form, which can be efficiently performed using the {\tt
BLAS} and {\tt LAPACK} (Anderson \etal 1999) libraries. All the
calculation parameters, such as the number of target states, continuum
orbitals, angular momenta, Floquet components as well as the
propagation distance, can be varied independently to ensure the
convergence of the results.

\section{Conclusion}

The $R$-matrix Floquet theory provides a powerful tool to study the
behaviour of a general atomic system in a laser field. The method is
fully {\sl ab initio} and non-perturbative and allows in particular an
exact treatment of all dynamic and radiative couplings. The general
approach for studying electron-atom scattering in a laser field is
similar to the treatment of multiphoton ionization presented in papers
I and II, but several aspects need more careful consideration, in
particular the boundary conditions. As we have shown in this paper, it
is necessary to define {\sl a priori} the field-dressed target states
to identify the collision channels and calculate cross sections. This
however has the advantage of greatly simplifying the formulae in the
outer region.

As in the standard $R$-matrix theory on which our method is based,
only one electron can be treated in the outer region. The energy of
the collisional electron and the field strength must not be too high,
so that ionization or photoionization of the target does not occur. 
The Floquet approach is furthermore only valid for a periodic field,
which again limits the field intensity as most powerful lasers are
pulsed.  Extensions of the $R$-matrix Floquet theory to treat stronger
fields are however possible. A time-dependent $R$-matrix theory has
already been proposed by Burke and Burke (1997) to handle non-periodic
fields in a one-dimensional model, which could be generalised to three
dimensions. There is furthermore no fundamental obstacle in combining
for instance the $R$-matrix pseudo-state method (Bartschat \etal 1996)
or the two-dimensional $R$-matrix propagator (Dunseath \etal 1996)
with the Floquet approach to include the double continuum. These
methods are computationally prohibitive at the moment but could become
realistic as the power of computers increases.

The theory presented in this paper has been successfully applied to
the study of free-free processes and simultaneous electron-photon
excitation (Charlo \etal 1998, Charlo 1999, Terao-Dunseath \etal
2001). These were the first accurate non-perturbative calculations
including a realistic multi-electron representation of the collision
system. We are currently continuing our study of electron-helium
scattering in a laser field (CO$_2$, Nd-YAG) and intend to apply the
method to other complex targets of experimental interest.

\ack
We would like to thank Maryvonne Le Dourneuf for many illuminating
discussions, Jean-Michel Launay for providing a number of useful
programs and Robert Allan for his help in parallelizing the new {\tt
RMF} computer package.

\References
\item[] Anderson E, Bai Z, Bischof C, Blackford S, Demmel J, Dongarra J,
Du Croz J, Greenbaum A, Hammarling S, McKenney A and Sorensen D
1999 {\em LAPACK Users' Guide} (Philadelphia, 
Society for Industrial and Applied Mathematics)
\item[] Andrick D and Langhans L 1976 \JPB {\bf 9} L459
\item[] \dash\ 1978 \JPB {\bf 11} 2355
\item[] Bader H 1986 \JPB {\bf 19} 2177
\item[] Bartschat K, Hudson E T, Scott M P, Burke P G and Burke V M
1996 \jpb {\bf 29} 115
\item[] Berrington K A, Burke P G, Butler K, Seaton M J, Storey P J,
Taylor K T and Yu Y 1987 \jpb {\bf 20} 6379
\item[] Berrington K A, Burke P G, Le Dourneuf M, Robb W D, Taylor K T 
and Vo Ky Lan 1978 {\em Comput. Phys. Commun.} {\bf 14} 367
\item[] Blackford L S, Choi J, Cleary A, D'Azevedo E, Demmel J, Dhillon I, 
Dongarra J, Hammarling S, Henry G, Petitet A, Stanley K, Walker D and
Whaley R C 1997 {\em ScaLAPACK Users' Guide} (Philadelphia, 
Society for Industrial and Applied Mathematics)
\item[] Bunkin F V and Fedorov M V 1966 {\em Sov. Phys. JETP} {\bf 22} 844
\item[] Burke P G and Berrington K A 1993 {\em Atomic and Molecular 
Processes: an $R$-matrix Approach} (Bristol: Institute of Physics)
\item[] Burke P G and Burke V M 1997 \jpb {\bf 30} L383
\item[] Burke P G, Burke V M and Dunseath K M 1994 \jpb {\bf 27} 5341 
\item[] Burke P G, Colgan J, Glass D H and Higgins K 2000 \jpb {\bf 33} 143
\item[] Burke P G, Francken P and Joachain C J 1990 {\em Europhys. Lett.} 
{\bf 13} 617
\item[] \dash\ 1991 \jpb {\bf 24} 761
\item[] Buttle P J 1967 \PR {\bf 160} 719
\item[] Charlo D, Terao-Dunseath M, Dunseath K M and Launay J-M 1998
\jpb {\bf 31} L539
\item[] Charlo D, 1999 {\em PhD thesis}, Universit\'e de Rennes 1 
(unpublished)
\item[] Colgan J, Glass D H, Higgins K and Burke P G 2001 \jpb {\bf 34}
2089
\item[] Dimou L and Faisal F H M 1987 \PRL {\bf 59} 872
\item[] D\"orr M, Terao-Dunseath M, Purvis J, Noble C J, Burke P G and
Joachain C J 1992 \jpb {\bf 25} 2809
\item[] D\"orr M, Terao-Dunseath M, Burke P G, Joachain C J, Noble C J
and Purvis J 1995 \jpb {\bf 28} 3545
\item[] Dunseath K M, Le Dourneuf M, Terao-Dunseath M and Launay J-M
1996 \PR A {\bf 54} 561
\item[] Eissner W, Jones M and Nussbaumer H 1974 {\em Comput. Phys. Comm.}
{\bf 8} 270
\item[] Ehlotzky F, Jaro\'n A and Kami\'nski J Z 1998 {\em Phys. Rep.}
{\bf 297} 63
\item[] Fainstein P D, Maquet A and Fon W C 1995 \jpb {\bf 28} 2723
\item[] Faisal F H M 1987 {\em Theory of Multiphoton Processes}
(New York and London: Plenum Press)
\item[] Francken P, Attaourti Y and Joachain C J 1988 \PR A {\bf 38} 1785
\item[] Geltman S and Maquet A 1989 \jpb {\bf 22} L419
\item[] Henneberger W C 1968 \PRL {\bf 21} 838
\item[] Hibbert A 1975 {\em Comput. Phys. Commun.} {\bf 9} 141
\item[] Jetzke S, Broad J and Maquet A 1987 \jpb {\bf 20} 2887
\item[] Jetzke S, Faisal F H M, Hippler R and Lutz H O 1984
\ZP A {\bf 315} 271
\item[] Joachain C J, D\"orr M and Kylstra N J 2000 {\em Adv. At. Mol. Opt. 
Phys.} {\bf 42} 225
\item[] Kramers H A 1956 {\em Collected Scientific Papers} (Amsterdam:
North-Holland) p 272
\item[] Kroll N M and Watson K M 1973 \PR A {\bf 8} 804
\item[] Kylstra N J and Joachain C J 1988 \PR A {\bf 58} R26
\item[] \dash\ 1999 \PR A {\bf 60} 2255
\item[] Luan S, Hippler R and Lutz H O 1991 \jpb {\bf 24} 3241
\item[] Manolopoulos D E 1986 \JCP {\bf 85} 6425
\item[] Mason N J 1993 \RPP {\bf 56} 1275
\item[] Mason N J and Newell W R 1987 \jpb {\bf 20} L323
\item[] \dash\ 1989 \jpb {\bf 22} 777
\item[] Mittleman M H 1993 \jpb {\bf 26} 2709
\item[] Rahman N K and Faisal F H M 1976 \JPB {\bf 9} L275
\item[] Shirley J H 1965 \PR B {\bf 138} 979
\item[] Smith F T 1969 \PR {\bf 179} 111 
\item[] Smith K 1971 {\em The Calculation of Atomic Collision Processes}
(New York: Wiley-Interscience) 
\item[] Terao-Dunseath M, Dunseath K M, Charlo D, Hibbert A and Allan R J
2001 \jpb {\bf 43} L263
\item[] Wallbank B, Connors V W, Holmes J K and Weingartshofer A 1987a
\JPB {\bf 20} L833
\item[] Wallbank B and Holmes J K 1993 \PR A {\bf 48} R2515
\item[] \dash\ 1994a \jpb {\bf 27} 1221
\item[] \dash\ 1994b \jpb {\bf 27} 5405
\item[] Wallbank B, Holmes J K and Weingartshofer A 1987b \JPB {\bf 20}
6121
\item[] \dash\ 1990 \jpb {\bf 23} 2997
\item[] Wallbank B, Holmes J K, LeBlanc L and Weingartshofer A 1988
\ZP D {\bf 10} 467
\item[] Wallbank B, Holmes J K, MacIsaac S C and Weingartshofer A 1992 
\jpb {\bf 20} 1265
\item[] Weingartshofer A, Clarke E M, Holmes J K and Jung C 1979 \PR A 
{\bf 19} 2371
\item[] Weingartshofer A, Holmes J K, Caudle G, Clarke E M and Kruger H 1977
\PRL {\bf 39} 269
\item[] Weingartshofer A, Holmes J K, Sabbagh J and Chin S L 1983 
\JPB {\bf 16} 1805
\endrefs

\end{document}